\documentclass[lettersize,journal]{IEEEtran}
\usepackage{cite}
\usepackage{amsmath,amssymb,amsfonts}
\usepackage{graphicx}
\usepackage{textcomp}
\usepackage{array}
\usepackage[caption=false,font=normalsize,labelfont=sf,textfont=sf]{subfig}
\usepackage[linesnumbered,ruled,vlined]{algorithm2e}
\usepackage{algpseudocode}
\usepackage{stfloats}
\usepackage{physics}
\usepackage{url}
\usepackage{caption}
\usepackage{manyfoot}
\usepackage{verbatim}
\usepackage{color,soul}
\usepackage{multirow}
\usepackage{subcaption}
\usepackage[colorlinks = true, citecolor = magenta]{hyperref}
\usepackage{mwe}
\usepackage{enumitem}
\usepackage{xcolor}
\usepackage{listings}
\usepackage{tabularray}
\usepackage{siunitx}

\begin{document}

\title{Low-latency machine learning FPGA accelerator for multi-qubit-state discrimination}

\author{%
	\IEEEauthorblockN{%
	    Pradeep~Kumar~Gautam,
            Shantharam~Kalipatnapu,
            Shankaranarayanan~H,   
            Ujjawal~Singhal,
            Benjamin~Lienhard,
		Vibhor~Singh and
		Chetan~Singh~Thakur 
	}\\

\thanks{

This work was supported by the Centre for Excellence in Quantum Technology (CoE-QT) through the Ministry of Electronics and Information Technology (MeitY) and Quantum Enabled Science and Technology (QuEST) Program within the Department of Science and Technology (DST), Government of India (GoI). The work of Vibhor Singh and Chetan Singh Thakur was supported by the Institute of Eminence (IoE) Scheme, Government of India.

Pradeep~Kumar~Gautam is with Department of Electronic Systems Engineering, Indian Institute of Science, Bangalore 560012, India and Defence Research and Development Organisation, Bangalore 560093, India.

Shantharam~Kalipatnapu,  Shankaranarayanan~H and Chetan~Singh~Thakur (Corresponding Author) are with Department of Electronic Systems Engineering, Indian Institute of Science, Bangalore 560012, India. (E-mail: csthakur@iisc.ac.in)

Benjamin~Lienhard is with Department of Chemistry, Princeton University, Princeton, NJ 08544, USA and Department of Electrical and Computer Engineering, Princeton University, Princeton, NJ 08544, USA.

Ujjawal~Singhal and Vibhor~Singh (Corresponding Author) are with the Department of Physics, Indian Institute of Science, Bangalore 560012, India. (E-mail: vsingh@iisc.ac.in)

(Corresponding authors: Vibhor Singh; Chetan Singh Thakur)

}

}

\maketitle




\begin{abstract}
Measuring a qubit state is a fundamental yet error-prone operation in quantum computing. These errors can arise from various sources, such as crosstalk, spontaneous state transitions, and excitations caused by the readout pulse. Here, we utilize an integrated approach to deploy neural networks onto field-programmable gate arrays (FPGA). We demonstrate that implementing a fully connected neural network accelerator for multi-qubit readout is advantageous, balancing computational complexity with low latency requirements without significant loss in accuracy. The neural network is implemented by quantizing weights, activation functions, and inputs. The hardware accelerator performs frequency-multiplexed readout of five superconducting qubits in less than \SI{50}{\nano\second} on a radio frequency system on chip (RFSoC) ZCU111 FPGA, marking the advent of RFSoC-based low-latency multi-qubit readout using neural networks. These modules can be implemented and integrated into existing quantum control and readout platforms, making the RFSoC ZCU111 ready for experimental deployment.

\end{abstract}


\maketitle

\section{Introduction}\label{sec:introduction}
\IEEEPARstart{Q}{uantum} processors are expected to solve specific computational tasks significantly more efficiently than their classical counterparts~\cite{boixo2018characterizing,shor1999polynomial,grover1996fast,schuld2015introduction,o2016scalable}. However, errors are inevitable during the operation of quantum processors due to the inherent instabilities of qubits and the challenges in controlling and reading out their quantum states, especially as quantum processors scale up in the number of qubits~\cite{barends2014superconducting}.

Quantum error correction (QEC) schemes address errors by redundantly encoding quantum information and performing repeated measurements to detect and correct errors during computation~\cite{lidar2013quantum}. The successful execution of QEC algorithms relies on constantly monitoring a group of qubits and swiftly invoking corrective action~\cite{google2021exponential}. To reduce resource requirements, especially from a readout perspective, multi-qubit readout is often achieved using frequency-division multiplexing~\cite{fowler2012surface}. Various methods, including matched filters, support vector machines (SVM), and neural networks (NN), have been used to process these frequency-multiplexed signals for inferring qubit states~\cite{ryan_tomography_2015,magesan2015machine,navarathna2021neural,lienhard2022deep,maurya2023scaling}.


When implementing such methods, minimizing latency in post-processing for qubit-state inference is crucial for enabling on-the-fly corrective actions in QEC schemes. NN-based state discriminators outperform traditional signal processing techniques in post-processing readout signals, as demonstrated on various qubit platforms~\cite{seif2018machine,ding2019fast,matsumoto2021noise,lienhard2022deep}. Additionally, NN-based discriminators scale efficiently, as they do not require qubit-specific processing, such as digital demodulation, for each qubit~\cite{lienhard2022deep}.

Scalable field-programmable gate array (FPGA) systems, particularly radio frequency system on chip (RFSoC) solutions for controlling and reading out individual qubits in quantum computing, have recently gained significant attention~\cite{stefanazzi2022qick,park2022icarus,tholen2022measurement,singhal2023sq}. In this context, NN-based state discriminators could significantly improve traditional signal processing solutions on FPGAs for frequency-multiplexed readout, enhancing throughput and reducing latency for real-time applications. However, a significant challenge is the substantial resource requirements for implementing NNs on hardware platforms like FPGAs~\cite{maurya2023scaling}. 

Overcoming these challenges can enhance readout signal post-processing, allow for additional error correction cycles in QEC protocols, and facilitate the integration of reinforcement learning agents into more complex systems~\cite{reuer2023realizing}. Quantization effectively reduces a model's storage and computational demands by representing parameters in low-precision fixed-point formats~\cite{gholami2022survey,nagel2021white}. Quantization-aware training (QAT) has been an effective method to quantize down model parameters to binary precision~\cite{blott2018finn,hubara2018quantized}.

In this work, we tackle the aforementioned challenges and demonstrate that we can design ultra-low latency NN-based qubit-state discriminators by employing QAT and automated flows for mapping NN onto an FPGA, we can design ultra-low latency, NN-based qubit-state discriminators. We present an integrated approach for an NN-based qubit-state discriminator design, covering everything from training to FPGA implementation, using tools such as Brevitas~\cite{brevitas} and FINN-R~\cite{blott2018finn}. This approach enables qubit-state inference with latencies ranging from \SI{24.03}{\nano\second} to \SI{47.12}{\nano\second}, significantly reducing the signal processing delays that have previously limited the number of QEC cycle repetitions.

The article is organized as follows: Section II covers the preliminaries. Section III details the design methodology and implementation of NN-based qubit-state discriminators. Section IV presents the results and compares them with state-of-the-art methods. Finally, Section V provides the conclusion of the work.

\begin{figure*}
\centering
\includegraphics[width=140mm]{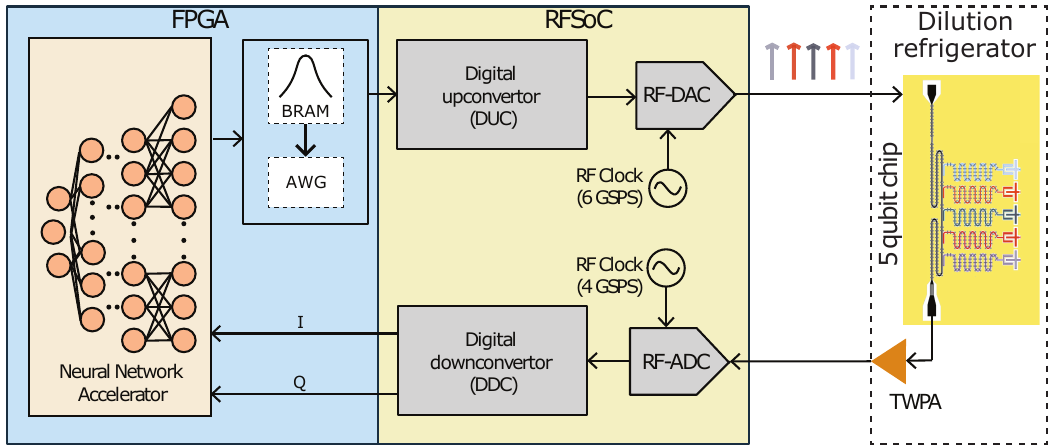}
\caption{Block diagram of a superconducting quantum processor interfacing with an FPGA system for qubit control and readout. The FPGA is part of the ZCU111 RFSoC evaluation kit. The quantum processor, which contains five superconducting qubits, is housed in a dilution refrigerator (for more details on the quantum processor and experimental setup, consult Ref.~\cite{lienhard2022deep}). RF digital-to-analog converter (RF-DAC) process control signals for the qubits. The readout signal combines all five readout tones and is transmitted through a single feedline. This signal is first amplified by, among others, a traveling-wave parametric amplifier (TWPA) and then digitized by an RF analog-to-digital converter (RF-ADC). The RF-ADC converts the frequency-multiplexed readout signal into in-phase (\(I\)) and quadrature (\(Q\)) components. A machine learning accelerator for qubit-state discrimination subsequently processes these components. Based on the inferred qubit states, a feedback signal may be generated to drive the subsequent control pulse generation logic.}
\label{fig:typical_setup}
\end{figure*}

\section{Superconducting Qubit Readout}

\begin{figure*}
\centering
\includegraphics[width=140mm]{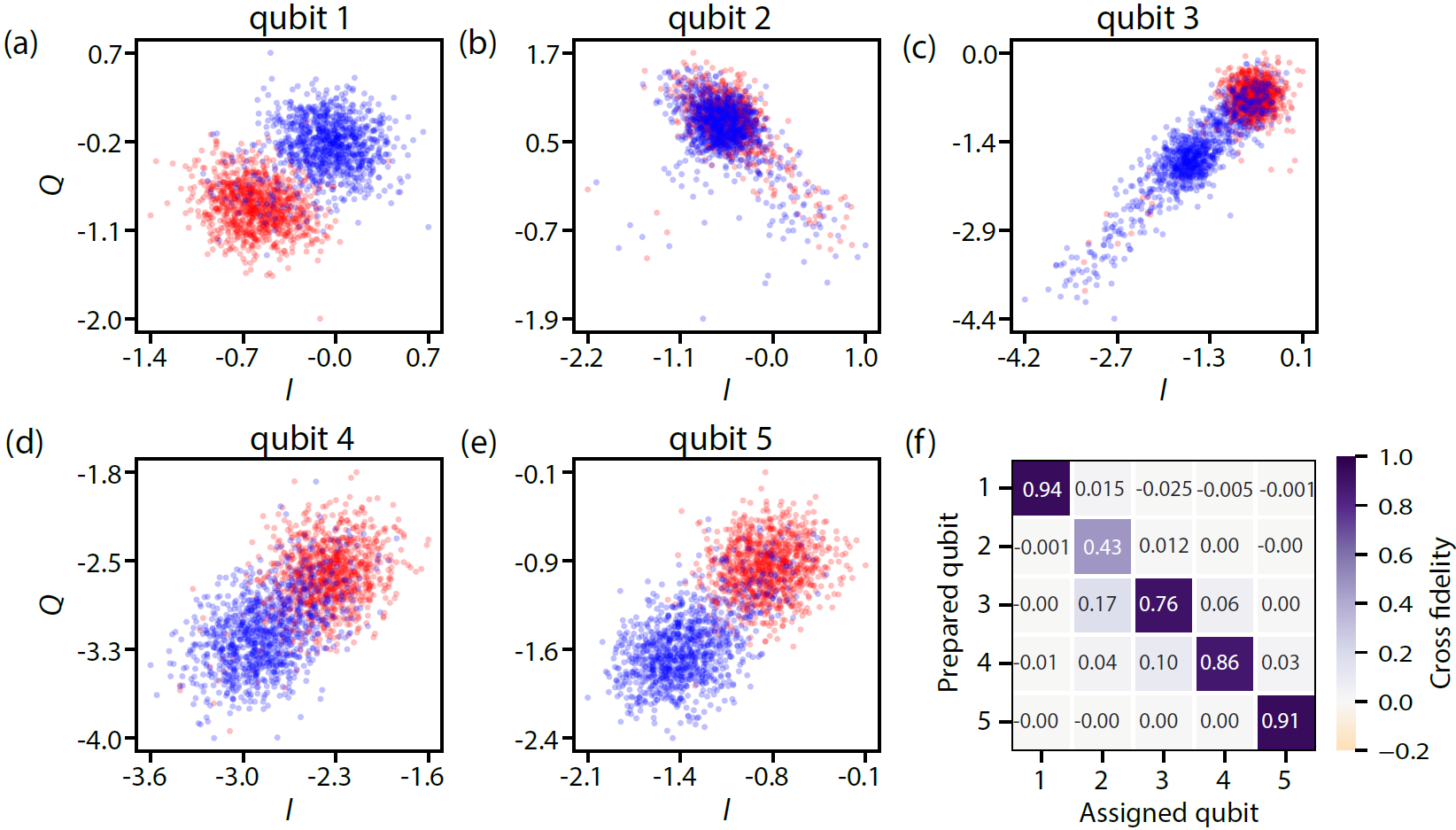}
\caption{Characteristics of qubit-specific single-shot readout traces. Panels (a-e) show the state discrimination of integrated single-shot traces for all five superconducting qubits, with red (blue) markers indicating the inferred ground (excited) state of each qubit. Panel (f) presents the cross-fidelity matrix, calculated using matched filters to infer the qubit states.}
\label{fig:single-shot}
\end{figure*}


A typical schematic of the experimental setup used to interface with a quantum processor comprising superconducting qubits is shown in Fig.~\ref{fig:typical_setup}. The setup includes a dilution refrigerator maintained at a temperature of around \SI{20}{\milli\kelvin} housing a five-qubit quantum processor and an RFSoC board at room temperature used to generate microwave pulses for qubit control and readout and post-process readout signals. Specifically, the readout probe pulse is passed through a radio frequency (RF) digital-to-analog converter (RF-DAC) and, after interacting with the quantum processor to acquire a qubit-state-specific signal characteristic, is fed into a high-speed RF analog-to-digital converter (RF-ADC). The RF-ADC and RF-DAC form the interface between the analog signals to and from the quantum processor and the digital RFSoC processing elements. 

We use the SQ-CARS architecture developed on the Xilinx RFSoC ZCU111~\cite{singhal2023sq}. The RFSoC device XCZU28DR includes eight high-precision, low-power RF-DACs and RF-ADCs with maximum sampling rates of $6.554$~GSPS and $4.096$~GSPS, respectively. These data converters are configurable and integrated with the programmable logic resources of the RFSoC via AXI interfaces, which are standard for exchanging data between connected components.

The superconducting transmon qubits studied here have frequencies between \SIrange{4.3}{5.2}{\giga\hertz} and energy relaxation times from \SIrange{7}{40}{\micro\second}. Detailed characterization of the device can be found in Ref.~\cite{lienhard2022deep}.

For the readout of all qubits, a frequency-multiplexed pulse with a duration of around \SI{1}{\micro\second}, consisting of superimposed signals at intermediate frequencies (IF) of \SI{64.729}{\mega\hertz}, \SI{25.366}{\mega\hertz}, \SI{24.79}{\mega\hertz}, \SI{70.269}{\mega\hertz}, and \SI{127.282}{\mega\hertz}, is upconverted to the readout resonators' frequencies in the \si{\giga\hertz}-range using a local oscillator frequency of \SI{7.127}{\giga\hertz}. The same local oscillator is used for down-conversion of the readout pulse after acquiring a qubit-state-specific phase shift for further digitization at a sampling rate of \SI{500}{\mega\hertz} and post-processing, such as qubit-state discrimination. 

The performance metric used to analyze the multi-qubit-state discrimination performance is \(F_{\rm GM}\), the geometric mean fidelity of the five qubits. The fidelity for the \(i\)-th qubit, \(F_i\), and \(F_{\rm GM}\) are defined as:
\begin{equation}
F_i = 1 - \frac{[P(0_i|\pi_i) + P(1_i|\emptyset_i)]}{2}
\label{fd equation}
\end{equation}
\begin{equation}
F_{\rm GM} = \sqrt[5]{F_1 F_2 F_3 F_4 F_5},
\label{fgm equation}
\end{equation}
where \(P(0_i|\pi_i)\) and \(P(1_i|\emptyset_i)\) represent the conditional probabilities of assigning the ground (excited) state with label $0$ ($1$) to qubit \(i\) when it is prepared in the excited (ground) state.

Multi-qubit processors often face the challenge of crosstalk, where signals---for control and readout---interfere with one another. Cross-fidelity is a key metric for assessing the performance of multi-qubit state discriminators amidst crosstalk, which quantifies the correlation between the assignment fidelities of individual qubits~\cite{PhysRevApplied.10.034040}. The cross-fidelity between two qubits is defined as:
\begin{equation}
    \label{cross_fd_eq}
    F_{ij}^{CF} = 1 - [P(1_i|\emptyset_j) + P(0_i|\pi_j)],
\end{equation}
where \(P(1_i|\emptyset_j)\) (\(P(0_i|\pi_j)\)) represents the preparation of qubit \(j\) in the ground (excited) state and subsequent detection of qubit \(i\) in the excited (ground) state. Positive (negative) off-diagonal elements represent a correlation (anti-correlation) between qubits. Here, in the experimental data, the observed correlations are primarily caused by readout crosstalk~\cite{lienhard2022deep}.

Fig.~\ref{fig:single-shot}(a)-(e) displays the results of integrated single-shot measurements in the \(IQ\)-plane for all five superconducting qubits. The cross-fidelity based on the matched filter state discriminator is shown in Fig.~\ref{fig:single-shot}(f). The off-diagonal colored cells indicate the correlation strength.

\section{Low-Latency Neural Network State Discriminator}

NN-based qubit-state discriminators have proven effective in mitigating readout crosstalk in multi-qubit systems~\cite{lienhard2022deep}. However, their integration onto dedicated hardware, such as RFSoCs, has not been accomplished yet due to the excessive storage and computational demands associated with these discriminators~\cite{maurya2023scaling}.

Quantization effectively reduces the storage and computational demands of models by representing parameters in low-precision fixed-point formats, often with minimal impact on accuracy~\cite{gholami2022survey,nagel2021white}. When employing QAT, low-precision quantized models, including those with binary precision, generally maintain high accuracy~\cite{blott2018finn, hubara2018quantized}. We utilize Brevitas' QAT~\cite{brevitas} to leverage arbitrary bit-width and mixed precision within the PyTorch framework~\cite{pytorch}.

\subsection{Neural Network Model Optimization and Quantization}

The potential for mixed-precision quantization of inputs, weights, and activations down to one bit within a large NN design necessitates careful architectural choices. To streamline design exploration, we adopted a strategy similar to that in Ref.~\cite{sung2015resiliency} and started with a base model presented in Ref.~\cite{lienhard2022deep}. This model has an input size of $1000$, which includes $500$ samples of each of the in-phase (\(I\)) and quadrature (\(Q\)) components corresponding to \SI{1}{\micro\second} of measurement time. The model features three hidden layers with $1000$, $500$, and $250$ nodes, respectively, and an output layer with $32$ nodes.

Initially, we adjusted the input feature size and the number of nodes in each hidden layer to the nearest powers of two. We then systematically reduced the input size and the number of nodes in the hidden layers. The number of output nodes is kept equal to the number of qubits rather than the total number of possible state combinations. This approach exponentially reduces the size of the output layer, using only five nodes (one for each qubit) instead of $32$ nodes, representing the \(2^N\) possible state combinations~\cite{lienhard2022deep, maurya2023scaling}. The training was conducted for $50$ epochs with a linearly decaying learning rate starting at \(10^{-3}\). 

The various model archetypes are represented as \( N_I \times N_{H_1} \times \dots \times N_{H_k} \times N_O \), where \( N_I \), \( N_{H_i} \), and \( N_O \) denote the number of nodes in the input layer, \( i^{th} \) hidden layer (\( i \in \mathbb{Z}^+ \)), and output layer, respectively.

Each hidden layer includes a linear transformation followed by batch normalization, a dropout layer, and a rectified linear unit (ReLU). The model inputs are derived from boxcar operations on \(I\) and \(Q\) samples by reducing the input feature size from $1024$ to $512$. We also examined the effect of model size on fidelity. Different architectures were trained using complete floating-point precision.

\begin{figure*}
\centering
\includegraphics[width=160mm]{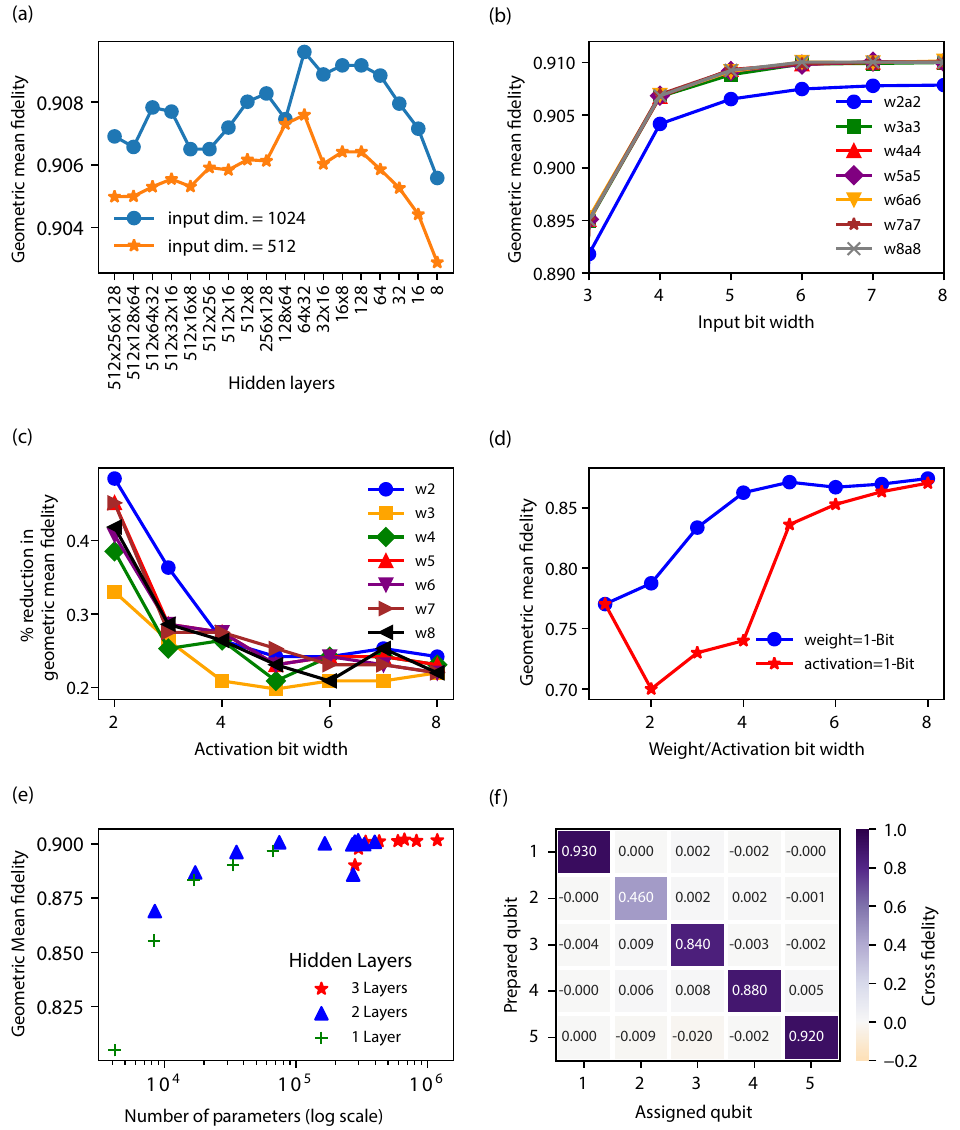}
\caption{Effects of neural network (NN) architecture parameters on readout fidelity. (a) Fidelity for various NN architectures with input feature sizes of $1024$ and $512$. The horizontal axis represents the dimensions of the hidden layers. (b) Impact on geometric mean fidelity \( F_{\rm GM} \) with varying input bit quantization sizes. The labels `w\#a\#' indicate the quantization of weights and activations where \# indicates the number of bits used for quantization. (c) Effects of mixed quantization of weights and activations on fidelity. (d) Impact of mixed quantization of weights and activations with binarized input. The blue (red) curve shows fidelity variation with activation (weight) bit width for weights quantized to a single bit. (e) Effect of model parameters and depth on readout fidelity. The input, weights, and activations are quantized to $4$, $2$, and $2$ bits, respectively. (f) Cross-fidelity matrix for the quantized NN. For panels (b), (c), (d), and (f), the NN architecture is \( 512 \times 64 \times 5 \).}
\label{FidelityVsParams}
\end{figure*}

Fig.~\ref{FidelityVsParams}(a) illustrates the combined effects of model size variation and input feature size on the geometric mean fidelity, \( F_{\rm GM} \). It appears that reducing the input feature size has a minimal impact on fidelity, a maximum of $0.3$\%. Similarly, decreasing the number of hidden layers to $2$ and a reduction in nodes composing the hidden layers only marginally affect \( F_{\rm GM} \). These results suggest that the model size can be carefully reduced without a significant readout fidelity loss. We selected the configuration \( 512 \times 64 \times 5 \), as further reduction in model size leads to a rapid decline in \( F_{\rm GM} \).

We perform quantization of the models to identify the optimal representation for input, weights, and activations. The training is conducted using the same hyperparameters as the earlier approach. We vary the bit-widths of input, weights, and activations from $8$-bit to $2$-bit and examine their effect on the \( F_{\rm GM} \), as shown in Fig.~\ref{FidelityVsParams}(b) for the model configuration \( 512 \times 64 \times 5 \). The results indicate that \( F_{\rm GM} \) degrades for input quantization below $4$-bit. Specifically, \( F_{\rm GM} \) drops to approximately $0.7$ for $2$-bit input quantization. This suggests that at least $4$-bit representation is necessary for accurate state discrimination in a five-qubit system. Additionally, the fidelity does not significantly vary with changes in input bit-width when activation and weight bit-widths are kept equal.

Fig.~\ref{FidelityVsParams}(c) displays the variation in \( F_{\rm GM} \) for mixed precision quantization of weights and activations, ranging from $8$-bit to $2$-bit. The results indicate that mixed precision does not significantly impact fidelity, consistent with the trend observed in Fig.~\ref{FidelityVsParams}(b). However, a model with binary quantization shows a notable decrease in accuracy, with fidelity dropping by $4$-$21$\%, as illustrated in Fig.~\ref{FidelityVsParams}(d). 

Fig.~\ref{FidelityVsParams}(e) explores the effect of network parameter count on \( F_{\rm GM} \) across different hidden layer sizes. It reveals that the fidelity remains relatively stable regardless of the hidden layer configuration when the number of parameters exceeds \( 10^4 \). Therefore, networks with fewer hidden layers are advisable to achieve lower latency while maintaining a high readout fidelity.

Based on the above-reported studies, we choose a quantization scheme of $4$-bit for the input and $2$-bit for both weights and activations. The resulting cross-fidelity matrix is computed and displayed in Fig.~\ref{FidelityVsParams}(f). This matrix indicates that the quantized neural network (QNN) significantly outperforms the matched filter discriminator and reduces crosstalk.

\subsection{FPGA Acceleration} 

Low-latency readout is a primary requirement for mid-circuit measurements and is essential to QEC. To achieve low-latency readout, the dataflow architecture-based framework of FINN-R~\cite{blott2018finn} stands out. FINN-R accepts models in the open NN exchange (ONNX) format with embedded FINN-specific metadata, which can be exported using the Brevitas library in the PyTorch environment. The model is then transformed into a streaming dataflow graph, with each node represented as a Xilinx high-level synthesis (HLS) callable function. Nodes without equivalent HLS functions would need to run on the processing system of the FPGA, resulting in increased latency.

FINN-R implements quantization and matrix multiplication of low-precision data using multi-vector threshold units (MVTU), with one such unit used for each layer of the QNN. Each unit consists of several processing elements with multiple input lanes, similar to a \textit{single instruction multiple data} architecture. The Appendix A provides more details on the FINN-R flow.

Trained and quantized models are adapted for varying levels of parallelization, constrained by resource availability, HLS conversion capabilities, and AXI-stream connection width. Layers that exceed a single MVTU unit are time-multiplexed, which impacts computation latency. A significant hurdle in achieving complete parallelism for moderate-size models is the limitation imposed by the connection width of the AXI-stream interconnect~\cite{pg035}. 

\begin{figure*}[!htb]
\centering
{\includegraphics[scale=0.78]{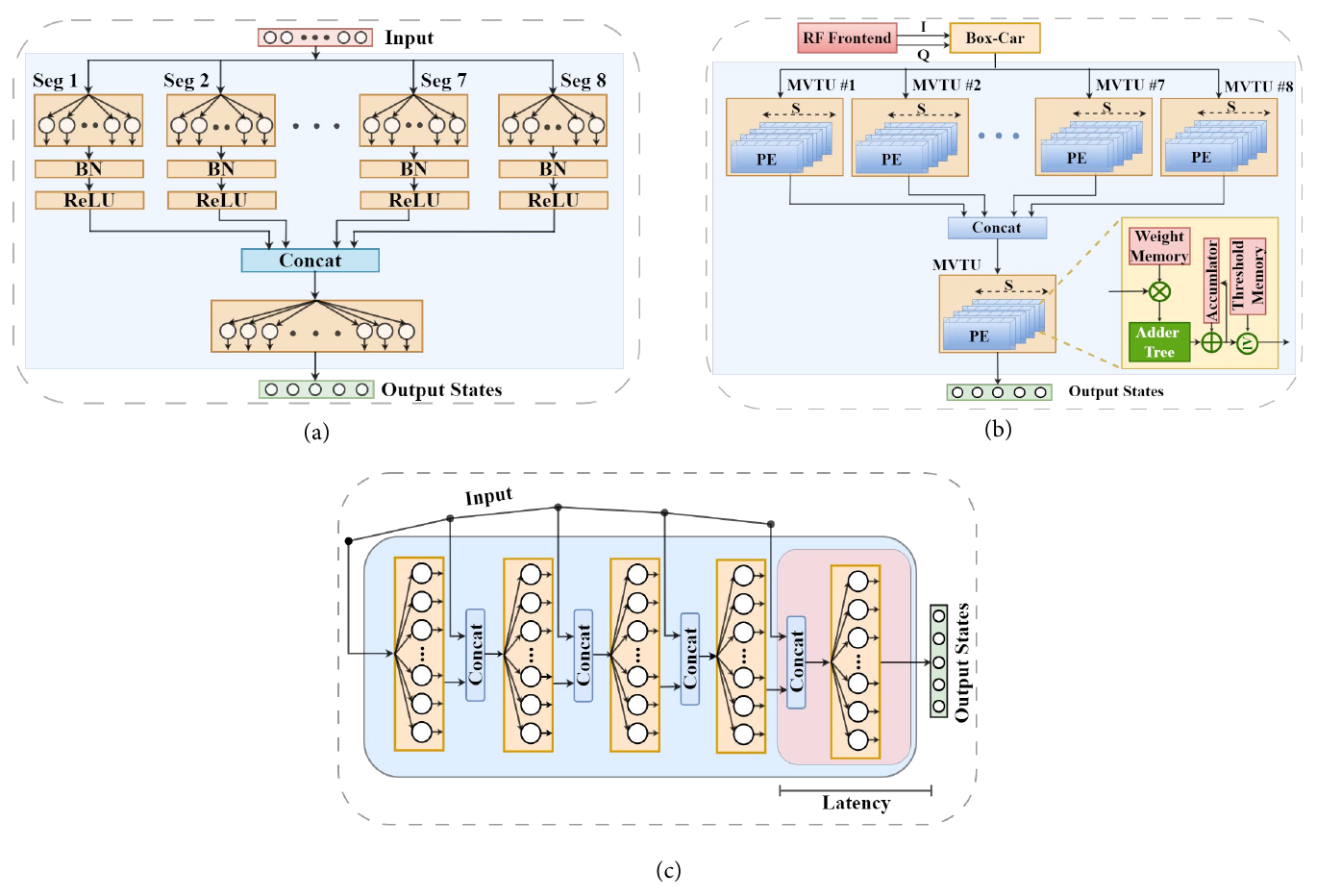}}
\caption{Quantized neural network (QNN) archetypes. (a) Software model of the QNN $\left(\left(512\times 8\right)\times 8\times 5\right)$ for achieving maximum parallelism. The first hidden layer consists of $8$ equal segments (Seg $1$ to Seg $8$), each containing eight nodes in the \textit{Linear} layer, followed by batch normalization (BN) and rectified linear units (ReLU). The output of each segment ($1\times 8$) is concatenated using the \textit{Concat} layer, resulting in a size of $1\times 64$. (b) Fully parallel hardware architecture of the QNN. Each segment of the model shown in panel (a) is implemented as a multi-vector threshold unit, which runs in parallel on hardware. A processing element, consisting of various computation blocks, is shown in the inset. (c) Piecewise layered QNN architecture $256\times 128\times 128\times 128\times 128\times 5$. Each layer processes a segment of the input signal along with the output from the previous layer. The input size for the first layer of the model is $1\times 256$, and subsequent layers receive an input size of $1\times 128$, composed of $1\times 64$ from the previous layer and $1\times 64$ from the input segment. The last section of the input signal is only fed to the last layer, meaning only the last layer contributes to the latency of the network.}
\label{Fig4:Architectures}
\end{figure*}

To address this constraint, we have designed a novel architecture to maximize the inherent parallelism of FPGAs. The modified hardware architecture is based on a NN with a configuration of $512 \times 64 \times 5$ as described earlier. The first hidden layer of the model, consisting of $64$ nodes, is divided into eight equal segments, each containing eight nodes. The $512$ nodes of the input layer are connected to all segments, effectively maintaining $64$ nodes in the first hidden layer. These segmented nodes can operate in parallel on the FPGA, eliminating the need for time-multiplexing and reducing total latency. As a result, the connection requirement for the first hidden layer decreases from $32,768$ to just $4,096$ per segment. The outputs of these segments are concatenated using a \textit{Concat} layer. The model architecture is illustrated in Fig.~\ref{Fig4:Architectures}(a). 

However, the generated ONNX model cannot be fully converted to a streaming dataflow-compliant representation using the default FINN-R flow due to the presence of the \textit{Concat} layer and non-uniform multiplication nodes preceding it. To achieve a fully dataflow-convertible model, we inserted uniform \textit{QuantIdentity} layers before the \textit{Concat} layer and introduced additional model graph modification steps within the default transformation process of the FINN-R flow. This novel architecture fully parallelizes the FPGA implementation of the model and reduces the latency by eliminating the need for time multiplexing. This approach effectively achieves ultra-low latency on FPGAs when working with large neural networks. Fig.~\ref{Fig4:Architectures}(b) displays the modified NN design generated by FINN-R.

As an alternative approach to attain low latency, we have also implemented a deeper NN that processes the input in a piecewise manner~\cite{reuer2023realizing}. In this configuration, only the last layer contributes to the latency of qubit-state discrimination, allowing for a deeper network with small hidden layers. The architecture for the model with a size of $256 \times 128 \times 128 \times 128 \times 128 \times 5$ is displayed in Fig.~\ref{Fig4:Architectures}(c).

\section{Results and Discussion}
\subsection{Performance and Resource Utilization}

To demonstrate the effectiveness of our approach, we implemented the NN model described in Ref.~\cite{lienhard2022deep}. This network architecture consists of an input dimension of $1,000$ nodes, three hidden layers with $1,000$, $500$, and $250$ nodes, and $32$ output nodes for each possible state. The total number of learnable parameters sums up to $1,634,782$, each requiring $4$ bytes of storage in floating-point representation. This translates to \SI{50}{\mega\byte} of storage just for the weights and biases. Using this standard floating-point approach, the implementation exceeds the resource limits of most available FPGA devices, as reported in Ref.~\cite{maurya2023scaling}.

To address this challenge, we adopted the methodology described in the previous section to implement the model in Brevitas and convert it to a dataflow graph using FINN-R. The input, weight, and activation quantization used are $4$, $2$, and $2$ bits, respectively. The model achieves a geometric mean fidelity of $0.904$, which is only $0.9$\% below the one reported in Ref.~\cite{lienhard2022deep}. This reduction is within the acceptable range reported by other quantized implementations of standard models~\cite{blott2018finn}.

Latency and resource utilization comparisons for various NN model architectures and quantizations are presented in Tab.~\ref{tab:resource_util_fpga}. Reducing input size and hidden layers have a significant impact on resource usage and drastically improve latency with a minimal drop in fidelity. 

The resource utilization of the base model (Arch-1) on the RFSoC XCZU28DR is only $39$\%, still leaving enough room for implementing other logic. Arch-4 to Arch-9 have almost the same parameters but vary in architecture and show a similar pattern in resource utilization, except for the binarized model of Arch-6, which uses less than about $5$\% of the look-up tables (LUT)---programmable logic blocks in FPGAs. Arch-7, Arch-8, and Arch-9 exhibit higher utilization as these architectures have been optimized for higher performance. 

The novel approach of splitting the first hidden layer of Arch-5 ($512 \times 64 \times 5$) into eight parallel segments results in Arch-7, reducing latency from $33$ cycles to $19$ cycles, an improvement of $42$\%. However, this comes with a significant increase in resource consumption, likely due to the additional \textit{QuantIdentity} layers introduced before the \textit{Concat} layer and the heightened level of parallelism. This architecture effectively balances complexity and performance. While binarized models are fast, they yield a significantly reduced fidelity. It is challenging to identify patterns in the maximum operable frequency of the design based on model architecture alone, as this is largely influenced by the Xilinx Vivado tool, which offers limited control over the outcome.

\begin{table*}[!htb]
\centering
\caption{Resource utilization and latency comparison of neural network archetypes.}
\begin{tabular}{|c|c|c|c|cc|c|c|c|}
\hline
                                                                                        &                                                                               &                                                                                   & \multicolumn{1}{l|}{}                           & \multicolumn{2}{c|}{Resource}                                                                                                                                                        &                                                                            &                                                                              &                                                                           \\ \cline{5-6}
\multirow{-2}{*}{Model Arch}                                                            & \multirow{-2}{*}{\begin{tabular}[c]{@{}c@{}}\# of \\ Parameters\end{tabular}} & \multirow{-2}{*}{\begin{tabular}[c]{@{}c@{}}Quantization\\ (IN/W/A)\end{tabular}} & \multicolumn{1}{l|}{\multirow{-2}{*}{$F_{\rm GM}$}} & \multicolumn{1}{c|}{\begin{tabular}[c]{@{}c@{}}LUT\\ (\% util)\end{tabular}}                         & \begin{tabular}[c]{@{}c@{}}FF\\ (\% util)\end{tabular}                        & \multirow{-2}{*}{\begin{tabular}[c]{@{}c@{}}Max Freq\\ (MHz)\end{tabular}} & \multirow{-2}{*}{\begin{tabular}[c]{@{}l@{}}Latency \\ (Cycles)\end{tabular}} & \multirow{-2}{*}{\begin{tabular}[c]{@{}c@{}}Latency \\ (ns)\end{tabular}} \\ \hline
\begin{tabular}[c]{@{}c@{}}$1000\times 1000\times 500\times 250\times 32$\\ (Arch-1)\end{tabular}                 & $1,634,782$                                                                     & $4/2/2$                                                                             & $90.40$                                           & \multicolumn{1}{c|}{\begin{tabular}[c]{@{}c@{}}$167054$\\ $\left(39.28\right)$\end{tabular}}                        & \begin{tabular}[c]{@{}c@{}}$135088$\\ $\left(15.88\right)$\end{tabular}                      & $243$                                                                        & $924$                                                                          & $3802.26$                                                                   \\ \hline
\begin{tabular}[c]{@{}c@{}}$1024\times 512\times 256\times 5$\\ (Arch-2)\end{tabular}                       & $657,413$                                                                       & $4/2/2$                                                                             & $90.39$                                           & \multicolumn{1}{c|}{\begin{tabular}[c]{@{}c@{}}$58336$\\ $\left(13.72\right)$\end{tabular}}                         & \begin{tabular}[c]{@{}c@{}}$46654$\\ $\left(5.48\right)$\end{tabular}                        & $275$                                                                        & $336$                                                                          & $1219.68$                                                                   \\ \hline
\begin{tabular}[c]{@{}c@{}}$512\times 128\times 32\times 5$\\ (Arch-3)\end{tabular}                         & $69,957$                                                                        & $4/2/2$                                                                             & $89.85$                                           & \multicolumn{1}{c|}{\begin{tabular}[c]{@{}c@{}}$38689$\\ $\left(9.10\right)$\end{tabular}}                          & \begin{tabular}[c]{@{}c@{}}$33358$\\ $\left(3.92\right)$\end{tabular}                        & $261$                                                                        & $51$                                                                           & $195.33$                                                                    \\ \hline
\begin{tabular}[c]{@{}c@{}}$512\times 64\times 32\times 5$\\ (Arch-4)\end{tabular}                          & $35077$                                                                         & $4/2/2$                                                                             & $90.11$                                           & \multicolumn{1}{c|}{\begin{tabular}[c]{@{}c@{}}$38087$\\ $\left(8.96\right)$\end{tabular}}                          & \begin{tabular}[c]{@{}c@{}}$46064$\\ $\left(5.41\right)$\end{tabular}                        & $350$                                                                        & $40$                                                                           & $114.4$                                                                     \\ \hline
\begin{tabular}[c]{@{}c@{}}$512\times 64\times 5$\\ (Arch-5)\end{tabular}                             & $33217$                                                                         & $4/2/2$                                                                             & $89.91$                                           & \multicolumn{1}{c|}{\begin{tabular}[c]{@{}c@{}}$43959$\\ $\left(10.37\right)$\end{tabular}}                         & \begin{tabular}[c]{@{}c@{}}$53252$\\ $\left(6.26\right)$\end{tabular}                        & $260$                                                                        & $33$                                                                           & $127.05$                                                                    \\ \hline
\begin{tabular}[c]{@{}c@{}}$512\times 64\times 5$\\ (Arch-6)\end{tabular}                             & $33217$                                                                         & $4/1/1$                                                                             & $80.1$                                            & \multicolumn{1}{c|}{\begin{tabular}[c]{@{}c@{}}$18302$\\ $\left(4.30\right)$\end{tabular}}                          & \begin{tabular}[c]{@{}c@{}}$15197$\\ $\left(1.78\right)$\end{tabular}                        & $440$                                                                        & $17$                                                                           & $38.59$                                                                     \\ \hline
{\color[HTML]{000000} \begin{tabular}[c]{@{}c@{}}$\left(\left(512\times 8\right)\times 8\right)\times 5$\\ \textbf{(Arch-7)}\end{tabular}} & {\color[HTML]{000000} $33295$}                                                  & {\color[HTML]{000000} $4/2/2$}                                                      & $89.72$                                           & \multicolumn{1}{c|}{{\color[HTML]{000000} \begin{tabular}[c]{@{}c@{}}$107026$\\ $\left(25.16\right)$\end{tabular}}} & {\color[HTML]{000000} \begin{tabular}[c]{@{}c@{}}$60696$\\ $\left(7.13\right)$\end{tabular}} & {\color[HTML]{000000} $403$}                                                 & {\color[HTML]{000000} $\mathbf{19}$}                                                    & {\color[HTML]{000000} $\mathbf{47.12}$}                                              \\ \hline
\begin{tabular}[c]{@{}c@{}}$256\times 128\times 128\times 128\times 128\times 5$\\ \textbf{(Arch-8)}\end{tabular}                  & $42124$                                                                         & $4/2/4$                                                                             & $89.78$                                           & \multicolumn{1}{c|}{\begin{tabular}[c]{@{}c@{}}$110351$\\ $\left(25.94\right)$\end{tabular}}                        & \begin{tabular}[c]{@{}c@{}}$123514$\\ $\left(14.52\right)$\end{tabular}                      & $341$                                                                        & $\mathbf{11}$                                                                           & $\mathbf{32.23}$                                                                     \\ \hline
\begin{tabular}[c]{@{}c@{}}$256\times 128\times 128\times 128\times 128\times 5$\\ \textbf{(Arch-9)}\end{tabular}                  & $42124$                                                                         & $4/1/4$                                                                             & $89.07$                                           & \multicolumn{1}{c|}{\begin{tabular}[c]{@{}c@{}}$104527$\\ $\left(24.58\right)$\end{tabular}}                        & \begin{tabular}[c]{@{}c@{}}$95945$\\ $\left(11.28\right)$\end{tabular}                      & $374$                                                                        & $\mathbf{9}$                                                                           & $\mathbf{24.03}$                                                                     \\ \hline
\end{tabular}
\label{tab:resource_util_fpga}
\
\end{table*}

Additionally, we have implemented an ultra-low latency SVM-based state discriminator on FPGA, which is particularly useful for single-qubit state discrimination. This method offers an improved decision boundary compared to a matched filter while maintaining an ultra-light hardware footprint. Further details on the SVM implementation can be found in the Appendix B.

\subsection{Comparison}

\begin{table*}[!htb]
\centering
\caption{Latency comparison of discriminator using neural networks with the state-of-the-art.}
\begin{tabular}{|c|c|c|c|c|c|}
\hline
                                                        & Discriminator                                                            & \# of Parameters & \begin{tabular}[c]{@{}c@{}}NN Latency\\ (ns)\end{tabular} & \begin{tabular}[c]{@{}c@{}}Resources\\ (LUTs)\end{tabular} & Readout Type \\ \hline
Reuer \textit{et al.}~\cite{reuer2023realizing} & $8$-point Boxcar Filter + NN                                                      & $1891$             & $48$                                                        & Not Reported                                               & Single       \\ \hline
Satvik \textit{et al.}~\cite{maurya2023scaling} & \begin{tabular}[c]{@{}c@{}}Demodulation + \\ Matched Filter + NN\end{tabular}        & $1112$             & Not Reported                                              & $17917$                                                      & Multiplexed  \\ \hline
\multirow{2}{*}{This Work}                              & \begin{tabular}[c]{@{}c@{}}$2$ point Boxcar Filter + NN \\ (Arch-7)\end{tabular} & $33295$            & $49.6$                                                        & $107026$                                                     & Multiplexed  \\ \cline{2-6} 
                                                        & \begin{tabular}[c]{@{}c@{}}$2$ Point Boxcar Filter + NN\\ (Arch-9)\end{tabular}  & $42124$            & $26.7$                                                     & $104527$                                                     & Multiplexed  \\ \hline
\end{tabular}
\label{tab:latencies}
\end{table*}

As an alternative to traditional signal processing, recent work has utilized NNs for their robustness to signal perturbations and their capability for multi-qubit state discrimination~\cite{lienhard2022deep,maurya2023scaling}. Maurya \textit{et al.}~\cite{maurya2023scaling} implemented a deep NN (DNN)-based discriminator for frequency-multiplexed qubit readout. Their approach uses a matched filter for dimensionality reduction before processing the data with a lightweight DNN. However, this approach introduces the additional requirement of qubit-specific demodulation and signal processing, which limits the solution's scalability. The latency for this approach is not reported and has $1,112$ learnable parameters.

Furthermore, Reuer \textit{et al.}~\cite{reuer2023realizing} describes a feed-forward NN designed for a reinforcement learning agent and state discriminator. This network features $7$ hidden layers, each with $20$ neurons, and is designed explicitly for single-qubit applications, generating decisions based on readout measurements. The approach has a latency of \SI{48}{\nano\second} and a total number of learnable parameters of $1,891$.

Tab.~\ref{tab:latencies} summarizes our networks' latency, resource usage, and number of learnable parameters compared to the state-of-the-art. Our NN qubit-state discriminator, based on Arch-7, Arch-8, and Arch-9 implemented on FPGA, has latencies of $20$ cycles, $12$ cycles, and $10$ cycles, respectively, including $1$ cycle for the boxcar operation to provide input to the NN. Consequently, Arch-7, Arch-8, and Arch-9 achieve latencies (including boxcar operation) of \SI{49.6}{\nano\second}, \SI{35.16}{\nano\second}, and \SI{26.7}{\nano\second}, respectively, which are comparable to or better than the state-of-the-art. Meanwhile, the presented approach has $18$ to $29$ times more learnable parameters, making it more robust and versatile for complex readout scenarios.

In Refs.~\cite{maurya2023scaling} and \cite{lienhard2022deep}, the number of output nodes corresponds to the total number of possible state combinations for $N$ qubits, $2^N$. However, using \textit{BCEWithLogitsLoss} as the loss function, the number of output nodes can be reduced to $N$. This reduction makes scaling the proposed NN qubit-state discriminator scalable in the number of qubits.

Our methodology is scalable and can support the implementation of large DNNs, as demonstrated by our implementation of the model reported in Ref.~\cite{lienhard2022deep} with $1.6$ million parameters. By employing QAT, we have shown that even with $2$-bit quantization, the model experiences only a minimal drop in fidelity while achieving ultra-low latency state discrimination, which is crucial for realizing quantum error correction (QEC), as evidenced by Arch-7, Arch-8, and Arch-9.

While Arch-7 effectively balances complexity and latency, some scenarios may require deeper networks. In such cases, Arch-8 and Arch-9 can achieve better latency, regardless of the network depth.

\section{Conclusion}
We present NN accelerators for state discrimination of frequency-multiplexed qubit readout traces of a superconducting multi-qubit processor. Our integrated approach deploys scalable NNs onto FPGAs, offering flexibility, automation, and faster design turnaround times. Adopting this methodology, we proposed ultra-low-latency architectures with latencies below \SI{50}{\nano\second}. The demonstrated substantial reduction in latency while maintaining qubit-state discrimination fidelity enables the implementation of QEC protocols with more error correction cycles, thereby significantly improving the performance of quantum processors. This marks the advent of low-latency NN architectures on FPGAs that do not require qubit-specific signal processing and can be scaled up as the number of qubits increases.

\appendices

\section{NN-Accelerator Design Methodology}

All models were trained and quantized using PyTorch 1.12.1 and Brevitas 0.9.1. The activation function employed is the rectified linear unit (\textit{ReLU}). We utilized the \textit{Adam}~\cite{Adam} optimizer with a weight decay of $10^{-3}$, a learning rate of $10^{-3}$, and a batch size of $1024$. During quantization-aware training (QAT), \textit{Linear} and \textit{ReLU} layers were replaced with their quantized equivalents, \textit{QuantLinear} and \textit{QuantReLU}, respectively.

\subsection{FINN-R Flow}

The conventional methods for implementing NNs on FPGAs typically involve either a hand-crafted custom architecture~\cite{reuer2023realizing} or high-level synthesis (HLS)~\cite{maurya2023scaling}. Both approaches restrict flexibility in selecting the NN's architecture and size. Additionally, custom architectures often entail long design turnaround times. Therefore, there is a need for an integrated approach that provides both automation and flexibility in implementing quantized neural networks (QNNs) on FPGAs.

\begin{figure}[!htb]
    \centering
    \includegraphics[scale=0.8]{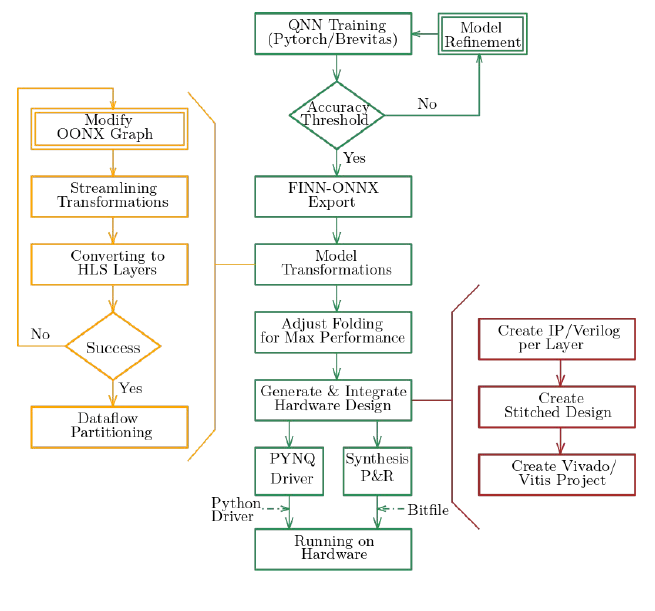}
    \caption{Flowchart illustrating the end-to-end process of FINN-R. The double-bordered rectangles indicate modifications made to the default FINN-R flow.}
    \label{fig:finn-flow-chart}
\end{figure}

Vitis-AI and FINN-R are leading automation frameworks for mapping QNNs onto FPGAs~\cite{hamanaka2023exploration}. Both frameworks use QNNs and fixed-point arithmetic to generate hardware designs, but they differ in their implementation strategies. Vitis-AI employs an overlay-based approach, which is highly scalable due to its use of off-chip memory for model parameter storage. However, this approach may not achieve the same performance levels as dataflow-based architectures. In contrast, FINN-R utilizes a dataflow architecture and relies on on-chip memory, resulting in lower latencies than overlay-based implementations. Given the critical need for low-latency readout in quantum technologies, the dataflow framework provided by FINN-R is the preferred choice for our work. The end-to-end flow of FINN-R is illustrated in Fig.~\ref{fig:finn-flow-chart}.

FINN-R processes models provided in open NN exchange format with FINN-specific metadata, which can be exported using the Brevitas library in a PyTorch environment. The framework transforms the model into a streaming dataflow graph and represents it in an intermediate representation. Nodes in the graph are then replaced with HLS-callable functions, and parallelism is either user-defined or derived based on throughput targets.

Quantization and matrix multiplication of low-precision data are handled by multi-vector threshold units (MVTU), with one unit per layer of the QNN. Each MVTU comprises multiple processing elements that operate similarly to single instruction multiple data architectures. Computation can be time-multiplexed to optimize hardware resource usage or dedicated processing elements and single instruction multiple data lanes can be employed for faster performance at the expense of higher hardware resource consumption. This flexibility allows for a balance between hardware resource usage and latency performance.

The final design is processed using the Xilinx Vivado back-end, which generates Verilog code for the targeted FPGA device. The design is exported as intellectual property, enabling the modification and integration into other designs based on specific requirements. The hardware architectures discussed are implemented on the Xilinx RFSoC ZCU111, with Xilinx Vivado 2022.2 used for design implementation. The synthesis strategy was optimized for high performance, with the primary goal of achieving low latency.

\section{SVM-based Qubit-State Discriminator}

Common single-qubit state discriminators include the boxcar~\cite{krantz2016single} and matched filter~\cite{turin1960introduction,ryan2015tomography,heinsoo2018rapid} with subsequent thresholding. Both methods are widely used for superconducting qubit platforms. A matched filter is generally preferred over a boxcar filter because it maximizes the signal-to-noise ratio. Moreover, boxcar filters are susceptible to additive stationary noise. Support vector machines (SVM) provide superior decision boundaries compared to boxcar or matched filters. Particularly, SVMs are more effective for state discrimination in single-qubit systems and frequency-division multiplexed readout, even when qubit-specific processing is involved~\cite{magesan2015machine,lienhard2022deep}.

The incoming stream of \(I\) and \(Q\) data from the RF-ADC is digitally demodulated at each qubit-specific intermediate frequency (IF) over the readout integration time. The training dataset consisted of $1,000$ random samples from each of $32$ possible combinations, with a vector size of $512$ for both \(IQ\)-data, corresponding to a \SI{1}{\micro\second} readout trace duration.

The linear support vector classification (LinearSVC) package~\cite{buitinck2013api} was utilized to implement the SVM. After training, we derived floating-point weights and biases, which were applied to the test data. The floating-point SVMs outperformed their matched filter counterparts, improving qubit fidelity and achieving a $1.53$\% increase in the overall geometric mean fidelity of the five qubits.

\begin{table}[!htb]
\centering
\caption{Geometric mean readout fidelities $F_{\rm GM}$ of all five qubits using matched filters and SVM discriminators.}
\begin{tabular}{|c|c|}
\hline
                                                                            & $F_{\rm GM}$  \\ \hline
Matched Filter (Float)            & $0.8846$  \\ \hline
SVM (Float)                                                                 & $0.8982$  \\ \hline
SVM (2 multiplier + 8-bit quant) & $0.8980 $ \\ \hline
SVM (1 multiplier + 8-bit quant) & $0.8985$  \\ \hline
\end{tabular}
\label{tab:svmfid}
\end{table}

\begin{figure}[!htb]
    \centering
    \includegraphics[scale=0.12]{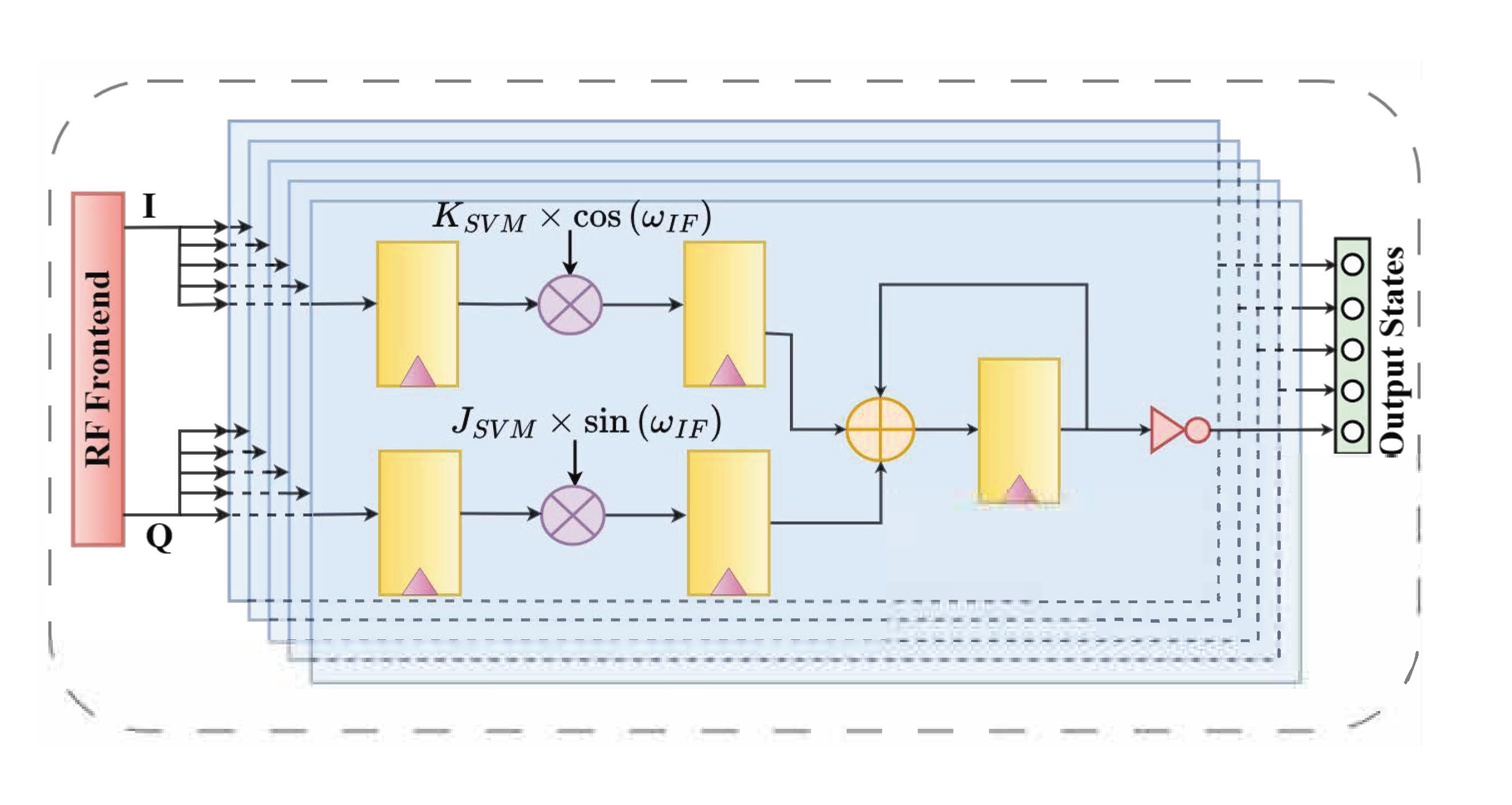}
    \caption{Signal processing for SVM-based qubit-state discriminator. The hardware architecture is designed to handle the incoming eight-bit \(I\) and \(Q\) inputs from the RF-ADC. It performs multiply-accumulate operations throughout the readout integration time. The resulting processing latency corresponds to the time required to process the final input sample. The architecture includes five SVM modules, one dedicated to each qubit.}
    \label{fig:svm_arch}
\end{figure}

\begin{table*}[!htb]
\centering
\caption{Latency comparison of SVM-based discriminator with the state-of-the-art.}
\begin{tabular}{|c|c|c|c|c|c|}
\hline
                                     & Discriminator      & \begin{tabular}[c]{@{}c@{}}Procesing Latency\\ Multi-cycle\\ (ns)\end{tabular} & \begin{tabular}[c]{@{}c@{}}Procesing Latency\\ Single-cycle\\ (ns)\end{tabular} & \# LUTs      & Readout type \\ \hline
Xiang \textit{et al.}~\cite{xiang2020simultaneous} & Demodulation + Boxcar Filter    & $32$                                                                             & Not reported                                                                    & Not Reported & Single       \\ \hline
Salathe \textit{et al.}~\cite{salathe2018low}     & Demodulation + Boxcar Filter  & $30$                                                                             & $5.3$                                                                             & $509$          & Single       \\ \hline
Yang \textit{et al.}~\cite{yang2022fpga}          & Demodulation + Boxcar Filter  & $20$                                                                             & Not Reported                                                                    & Not Reported & Single       \\ \hline
Guo \textit{et al.}~\cite{guo2022low}              & Demodulation + Matched Filter  & $24$                                                                             & Not Reported                                                                    & Not Reported & Single       \\ \hline
Thol{\'e}n \textit{et al.}~\cite{tholen2022measurement}      & Demodulation + Matched Filter  & $10$                                                                             & Not Reported                                                                    & Not Reported & Single       \\ \hline
This work                            & Demodulation + SVM & {$5.74$}                                                                           & {$3.67$}                                                                            & $1675$         & Multiplexed  \\ \hline
\end{tabular}
\label{lat:svm}
\end{table*}

The weights, along with the input and demodulation parameters, were quantized to eight bits. In the quantized implementation, we have two variants: one uses separate multipliers for demodulation and weight multiplication. At the same time, the other employs a standard multiplier that processes pre-computed fused values of weights and demodulation coefficients, reducing the number of multipliers by one per qubit. Tab.~\ref{tab:svmfid} compares the readout fidelities of this implementation with those of matched filters and various SVM approaches.

The hardware implementation of the SVM is illustrated in Fig.~\ref{fig:svm_arch}. The quantized SVM implementation, which optimizes weight and computation efficiency, outperforms other state-of-the-art traditional signal processing discriminators regarding latency and resource utilization. Tab.~\ref{lat:svm} details a comparison with other leading methods.

Salathe \textit{et al.}~\cite{salathe2018low} proposed a method using demodulation and thresholding, achieving a core processing latency of \SI{30}{\nano\second}. Tholen \textit{et al.}~\cite{tholen2022measurement} presented an integrated RFSoC solution with a matched filter state discriminator that uses pre-stored samples, resulting in a readout latency of \SI{10}{\nano\second}. Guo \textit{et al.}~\cite{guo2022low} demonstrated demodulation and matched filtering with a readout latency of \SI{24}{\nano\second}. In contrast, our quantized SVM on FPGA achieves a discriminator latency of \SI{5.74}{\nano\second} and utilizes only $1675$ LUTs for the five-qubit system, demonstrating significant performance and resource efficiency improvements.

\section*{Acknowledgment}
Ujjawal Singhal acknowledges the support received through the 
Prime Minister’s Research Fellowship (PMRF), GoI. Pradeep Kumar Gautam thanks Ankita Nandi and Vignesh Ramanathan for valuable suggestions on the manuscript.

\bibliographystyle{IEEEtran}
\footnotesize{
	\bibliography{ref}
}

\end{document}